\documentclass[aps,prl,reprint,superscriptaddress,amsmath,amssymb,nobibnotes]{revtex4-2}

\usepackage{amsmath}
\usepackage{amssymb}
\usepackage{amsfonts}
\usepackage{mathrsfs}
\usepackage{mathtools}
\usepackage{graphicx}
\usepackage{bm}
\usepackage{braket}
\usepackage{bbold}
\usepackage{multirow}
\usepackage{array}
\usepackage{makecell}
\usepackage{tabulary}
\usepackage{longtable}
\usepackage{float}
\usepackage{xspace}
\usepackage{pifont}
\usepackage{ulem}
\usepackage{xcolor}
\usepackage[colorlinks=true,linkcolor=blue,citecolor=blue,urlcolor=blue]{hyperref}
\definecolor{bule}{rgb}{0,0,1} 

\newcolumntype{C}[1]{>{\centering\arraybackslash}p{#1}}
\newcolumntype{L}[1]{>{\raggedright\arraybackslash}p{#1}}

\begin{document}
\title{Interstitial Magnetoelectricity in Altermagnetic Electrides}

\author{Yue Li}
\affiliation{School of Physics, Beihang University, Beijing 100191, China}
\affiliation{Science, Mathematics and Technology (SMT) Cluster, Singapore University of Technology and Design, Singapore 487372, Singapore}

\author{Shibo Fang}
\email{shibo\_fang@sutd.edu.sg}
\affiliation{Science, Mathematics and Technology (SMT) Cluster, Singapore University of Technology and Design, Singapore 487372, Singapore}

\author{Jianhua Wang}
\affiliation{School of Materials Science and Engineering, Tiangong University, Tianjin 300387, China}
\affiliation{Science, Mathematics and Technology (SMT) Cluster, Singapore University of Technology and Design, Singapore 487372, Singapore}

\author{Qiang Li}
\affiliation{School of Physics and Electronic Information Engineering, Hubei Minzu University, Enshi 445000, China}
\affiliation{Science, Mathematics and Technology (SMT) Cluster, Singapore University of Technology and Design, Singapore 487372, Singapore}

\author{Weizhen Meng}
\affiliation{College of Physics, Hebei Key Laboratory of Photophysics Research and Application, Hebei Normal University, Shijiazhuang 050024, China}

\author{Haiyu Meng}
\affiliation{School of Physics and Optoelectronics, Xiangtan University, Xiangtan 411105, China}

\author{Xiong-Xiong Xue}
\affiliation{School of Physics and Optoelectronics, Xiangtan University, Xiangtan 411105, China}

\author{Jianfeng Wang}
\affiliation{School of Physics, Beihang University, Beijing 100191, China}

\author{Yi Du}
\email{yi\_du@buaa.edu.cn}
\affiliation{School of Physics, Beihang University, Beijing 100191, China}

\author{Jing Lu}
\email{jinglu@pku.edu.cn}
\affiliation{State Key Laboratory for Mesoscopic Physics and School of Physics, Peking University, Beijing 100871, China}

\author{Yee Sin Ang}
\email{yeesin\_ang@sutd.edu.sg}
\affiliation{Science, Mathematics and Technology (SMT) Cluster, Singapore University of Technology and Design, Singapore 487372, Singapore}

\begin{abstract}
Electrides and altermagnets are two distinct yet intriguing classes of quantum materials, characterized by interstitial anionic electrons and nonrelativistic spin splitting, respectively. The recent prediction of Stoner-instability-driven altermagnetic electride provides a natural bridge between these two fields. Here, we investigate the magnetoelectric response and its related spin-polarized transport in the altermagnetic electride Zr$_2$N monolayer. Using the tight-binding Hubbard model and first-principles calculations, we find that the electric field can break the magnetic compensation between the two interstitial anionic electrons, inducing a finite net magnetization and giving rise to an interstitial electron mediated magnetoelectric response. The magnetoelectric response is further demonstrated through the electrical control of spin conductivity. Our results reveal a magnetoelectric effect in altermagnetic electrides, paving a way towards electride-based spintronic devices.
\end{abstract}


\maketitle


\section{Introduction}
Electride is an intriguing class of materials in which excess electrons occupy interstitial sites and act as anions, giving rise to interstitial anionic electrons (IAEs) with unusual electronic structures and spatially extended charge distributions~\cite{HosonoKitano2021Electrides,Lim2024,6dv1-7ddd,Geng2026,PRL2026}. Owing to these characteristics, electrides have attracted broad interest in various fields, including superconductivity~\cite{Miyakawa2007ElectrideSC,Zhang2017}, high-pressure physics~\cite{Ma2009DenseSodium}, topology~\cite{mengAM2025,PhysRevLett.125.187203}, catalysis~\cite{Kitano2012Ammonia,Kitano2015}, and negative-refraction optics~\cite{Guan2017NegativeRefraction}. Beyond experimentally realized electrides such as
Ca$_2$N, Gd$_2$C, and YCl~\cite{Lee2013Ca2N,Lim2024,Geng2026}, high-throughput studies have further expanded the landscape of inorganic electrides~\cite{Burton2018ElectrideScreening,Meng2024ElectrideReview} and identified diverse magnetic-electride candidates~\cite{Zhang2023MagneticElectrides,Meng2025MagneticElectrideReview}. 
In parallel, altermagnetism has emerged as a distinct form of collinear magnetism that combines vanishing net magnetization with momentum-dependent spin splitting~\cite{Smejkal2022Beyond,Smejkal2022Landscape,Krempasky2024Altermagnetic,PhysRevLett.133.206401-2024,PhysRevLett.132.036702-2024,Jungwirth2026Altermagnetism,Yu2026,Li2026}, offering new opportunities for spintronic applications~\cite{Zhu2024,Zhou2025,PhysRevLett.134.106801,PhysRevLett.134.106802-liu,Wang2026Pentagonal,Fang2026EdgeStates}.
Recently, these two material concepts are brought together through the proposal of electride altermagnets, with monolayer Zr$_2$N and Ti$_2$N identified as representative realizations~\cite{Cheng2026InterstitialAM}. In these systems, altermagnetism originates from the interplay among the Stoner instability of IAEs, superexchange interactions, and the underlying spin-space-group symmetry. The resulting systems exhibit pronounced piezo-magnetic responses and unconventional ultrafast excitation dynamics, establishing electride altermagnets as an intriguing platform for spintronic applications.

Magnetoelectricity provides a pathway for controlling magnetic properties through electric fields~\cite{Lottermoser2004,Eerenstein2006}. Magnetoelectric responses have been extensively investigated in multiferroic materials~\cite{10.1021/jacs.5c16402,peng2025ferroelastic} and magnetic heterostructures~\cite{Rondinelli2008,PhysRevLett.107.217202,xuexiongxiongPRB}, involving mechanisms such as spin–orbit coupling, exchange interactions, and lattice-mediated coupling~\cite{Eerenstein2006,Eerenstein2007,PhysRevLett.104.127202,Yang2014,Spaldin2019}. Recent studies have proposed gate-controlled magnetoelectricity of altermagnets through spin-layer coupling~\cite{Zhang2024GateField,Peng2025LayerSpin}. In electrides, interstitial electrons can possess intrinsic magnetic moments, as experimentally established in layered Gd$_2$C and mixed-cation electrides~\cite{Lee2020MagneticElectride,Lee2021CavityFerrimagnetism}. Their spin polarization depends sensitively on the local cavity environment and electron occupancy, as further evidenced by reversible magnetic reconstruction through electron--anion substitution in electride--hydride systems~\cite{Lee2023ElectrideHydride}. These observations point to the interstitial-electron charge distribution as an additional degree of freedom for controlling magnetism, without necessarily relying on conventional atom-centered magnetic moments.

Here, we establish an interstitial electron mediated magnetoelectric response as a mechanism for the electrical control of IAEs altermagnetism, using monolayer Zr$_2$N as a representative example. Combining the tight-binding (TB) Hubbard model with first-principles calculations, we investigate the electronic structures, spin transport properties, and electron redistribution under an out-of-plane electric field. 
We show that the electric field modifies the spin polarization of the interstitial electrons, driving a transition from altermagnetism to ferrimagnetism. Furthermore, the electrically induced magnetic reconstruction produces tunable spin-polarized transport. These findings extend the scope of magnetoelectric responses beyond conventional atom-centered magnetism, unraveling altermagnetic electrides as a promising platform for the electrical control of magnetism and spin transport.

\begin{figure}[t]
\centering
\includegraphics[width=1\linewidth]{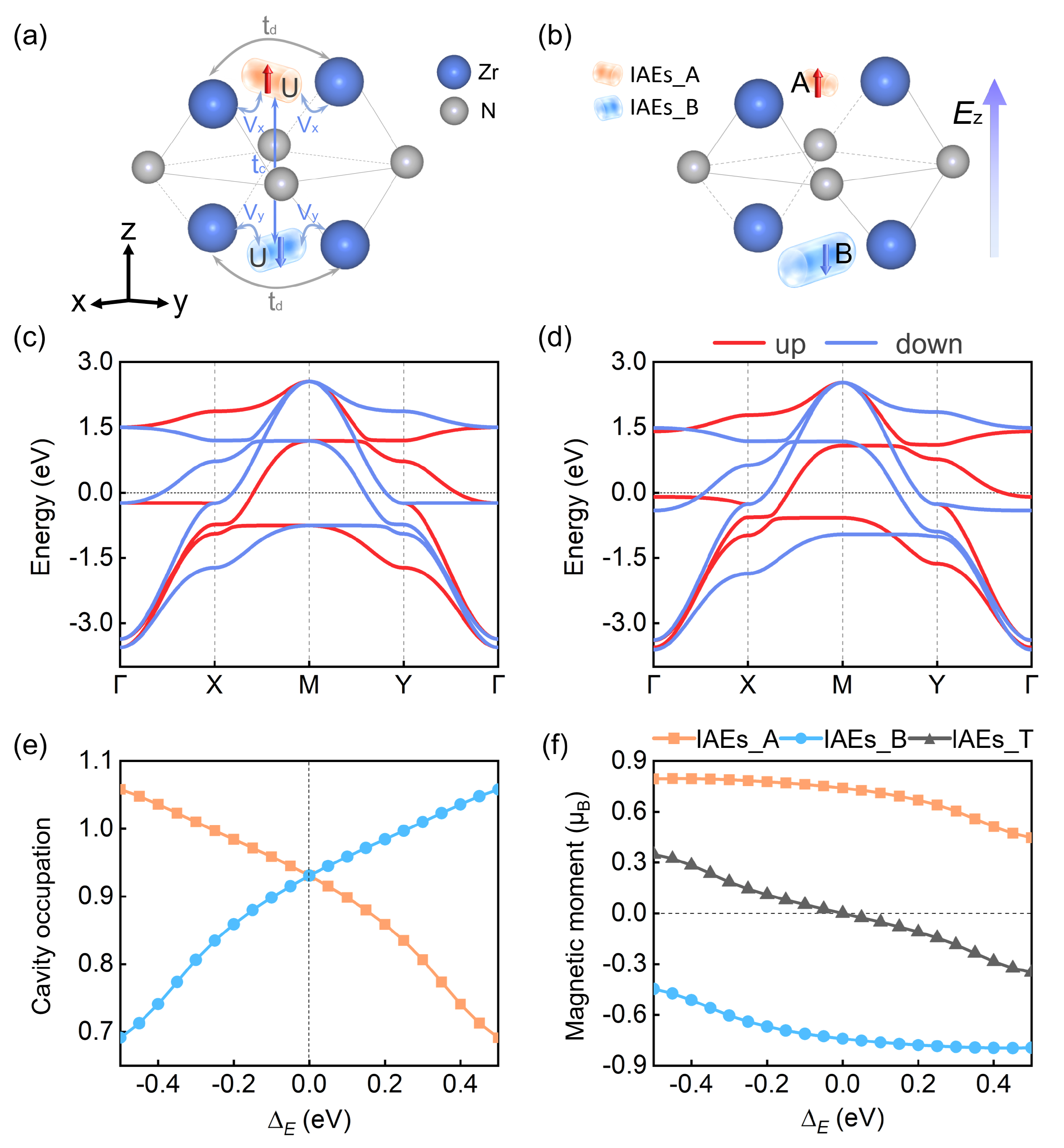}
\caption{\textbf{Tight-binding Hubbard model of monolayer Zr$_2$N.}
(a, b) Model schematics and (c, d) corresponding spin-resolved band structures at $E_z=0$ and $E_z\neq 0$, respectively. 
(e) The cavity occupations and (f) magnetic moments for the different electrostatic shifts $\Delta_E$.  
\label{Figure1}}
\end{figure}

\section{Results}

\subsection{TB model of altermagnetic electride Zr$_2$N}

The interstitial magnetoelectric response refers to the coupling between an electric field and magnetism carried by interstitial electrons~\cite{Sui2021}. To investigate this response, we consider the altermagnetic electride monolayer Zr$_2$N. The magnetism of monolayer Zr$_2$N is predominantly carried by IAEs confined in two cavities, rather than by atom-centered magnetic orbitals~\cite{Cheng2026InterstitialAM}. We construct a four-orbital TB Hubbard model comprising two IAEs cavity orbitals ($c_A$ and $c_B$) and their hybridized Zr-$d$ states ($d_A$ and $d_B$), as illustrated in Figs.~\ref{Figure1}(a) and (b). Here, $\mathrm{IAEs\_A}$ and $\mathrm{IAEs\_B}$ denote the IAEs in cavities A and B, respectively, as shown in Fig.~\ref{Figure1}(b). The Hamiltonian reads
\begin{equation}
\begin{aligned}
H={}&\sum_{\mathbf{k}\sigma}
\Psi_{\mathbf{k}\sigma}^{\dagger}
h_0(\mathbf{k})\Psi_{\mathbf{k}\sigma}\\
&+U\sum_{i,\alpha}n_{i\alpha\uparrow}n_{i\alpha\downarrow}
+\Delta_E\sum_i(n_{iA}-n_{iB}),
\end{aligned}
\label{eq:spin_cavity}
\end{equation}
where $\Psi_{\mathbf{k}\sigma}=
(d_{A\mathbf{k}\sigma},d_{B\mathbf{k}\sigma},
c_{A\mathbf{k}\sigma},c_{B\mathbf{k}\sigma})^T$
and $\alpha=A,B$ labels the two cavities.
The Hamiltonian $h_0(\mathbf{k})$ incorporates the Zr-$d$ and cavity onsite energies ($\varepsilon_d$ and $\varepsilon_c$), their hopping amplitudes ($t_d$ and $t_c$), and the directional Zr-IAEs hybridizations
$V_x(\mathbf{k})=2V\cos(k_xa/2)$ and
$V_y(\mathbf{k})=2V\cos(k_ya/2)$.
The inequivalent hybridization channels generate a cavity self-energy difference proportional to
$\cos(k_xa)-\cos(k_ya)$, giving rise to the characteristic $d$-wave momentum dependence of the nonrelativistic spin splitting, consistent with the previous density functional theory (DFT) results~\cite{Cheng2026InterstitialAM}.

The onsite Hubbard interaction $U$ on each cavity drives the Stoner instability of the IAEs, while the cavity hopping $t_c$ generates an antiferromagnetic superexchange of order $t_c^2/U$ in the strong coupling limit. We treat the Hubbard interaction self-consistently within the collinear mean-field approximation~\cite{Hubbard1963,Bach1994}, obtaining an altermagnetic (AM) state energetically favored over the ferromagnetic (FM) and nonmagnetic (NM) configurations at zero field~\cite{PhysRevB.37.9753}. We obtain the band structure under zero field which exhibits nonrelativistic spin splitting along the $X$--$M$--$Y$ path in Fig.~\ref{Figure1}(c). The model details and parameters are provided in the Supporting Information.

We further investigate the magnetoelectric response under an out-of-plane, perpendicular electric field.
The perpendicular electric field introduces opposite electrostatic shifts $\pm\Delta_E$ to the two cavities, where $\Delta_E=eE_zd_{\rm eff}/2$ and $d_{\rm eff}$ denotes the effective distance between the two cavities. As shown in Fig.~\ref{Figure1}(d), the band degeneracy under zero field is lifted, accompanied by a transition from altermagnetism to ferrimagnetism in the presence of the electric field. The electron occupations in the upper and lower cavities become unequal, resulting in unequal magnetic moments and a net spin polarization in Figs.~\ref{Figure1}(e) and (f). Upon reversing the electric field, the electron occupations satisfy
$n_A(E_z)=n_B(-E_z)$, while the magnetic moments of IAEs satisfy 
$M_A(E_z)=-M_B(-E_z)$. Moreover, the net spin polarization ($\mathrm{IAEs\_T}$) gradually increases with increasing electric field. The explicit expressions are provided in the Supporting Information. Consequently, the electric field induces a magnetoelectric response by breaking the balance of electron occupations and the net spin polarization between the two cavities.

\subsection{DFT calculation of altermagnetic electride Zr$_2$N}

\begin{figure}
\centering
\includegraphics[width=\columnwidth]{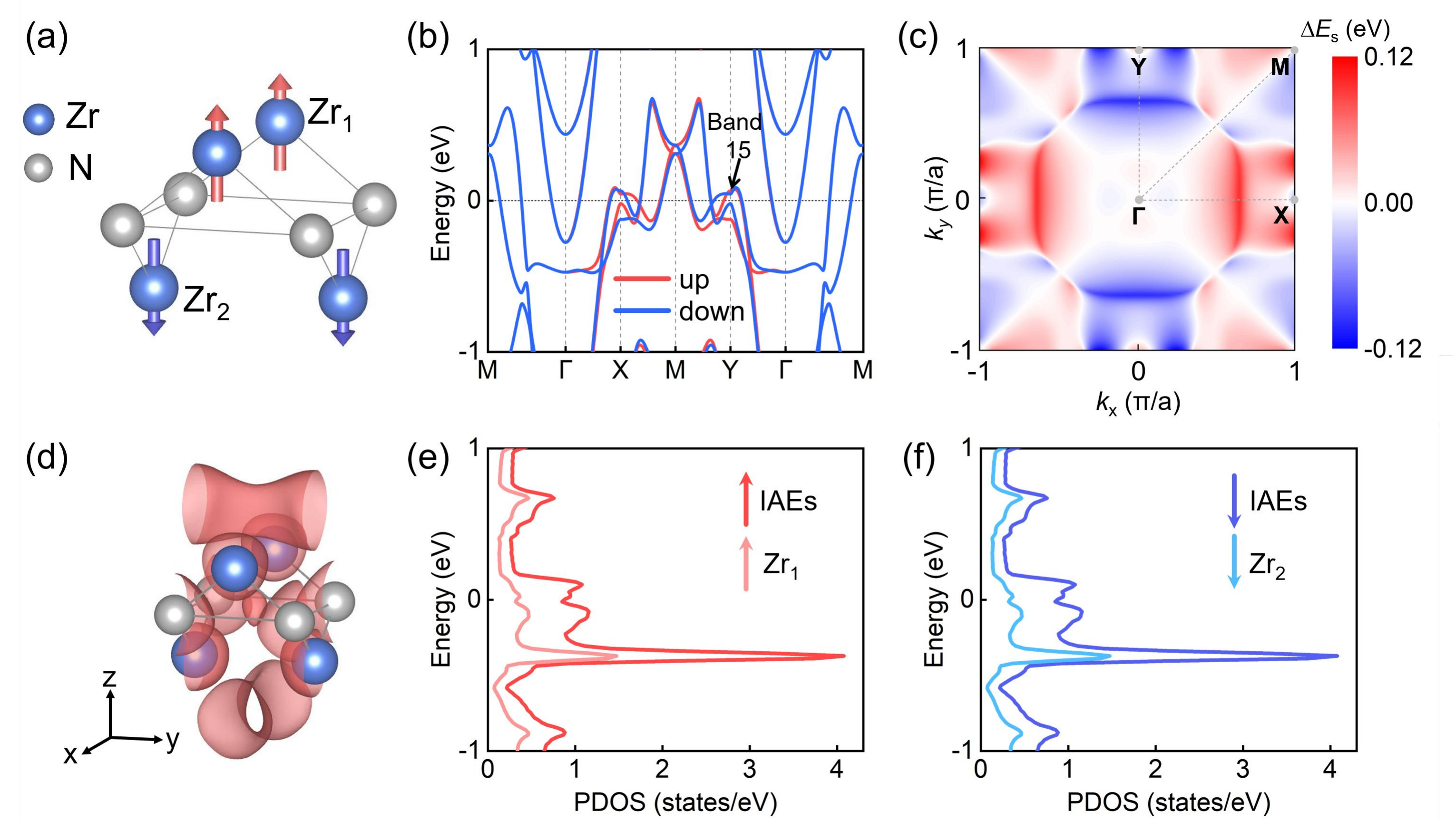}
\caption{\textbf{Interstitial-electron altermagnetism in monolayer Zr$_2$N.}
(a) Crystal and magnetic structure.
(b) Spin-resolved band structure.
(c) Momentum-resolved $d$-wave spin splitting of band 15, $\Delta E_s(\mathbf{k})=E_{\uparrow}(\mathbf{k})-E_{\downarrow}(\mathbf{k})$.
(d) Electron localization function (ELF). The isosurface value is set as 0.3.
(e, f) Projected density of states (PDOS) of IAEs and Zr atoms for spin-up and spin-down channels, respectively.
\label{Figure2}}
\end{figure}

Monolayer Zr$_2$N was initially identified as a two-dimensional cross-chain electride, characterized by spatially separated interstitial electron channels, and was predicted to exhibit superconductivity with a transition temperature of $T_c=0.6$ K~\cite{Zha2025CrossChainElectride}. The spin space group of the altermagnetic state is $P^{\bar{1}}\bar{4}^{1}m^{\bar{1}}2^{\infty 100 m}1$ (115.25.1.1.L)~\cite{Yu2026FINDSPINGROUP}. The two spin-opposite IAEs sublattices are connected by the combined spin-space symmetry $[C_2||S_4]$, making the system an altermagnet~\cite{Cheng2026InterstitialAM}. 

The structural and electronic properties of monolayer Zr$_2$N are shown in Fig.~\ref{Figure2}. The monolayer adopts a modified Lieb-lattice geometry, with two Zr atoms located in the upper and lower atomic layers and one N atom in the central layer, as shown in Fig.~\ref{Figure2}(a). The optimized lattice constants are $a=b=3.24$~\AA. The band structure reveals metallic behavior with pronounced nonrelativistic spin splitting along the $X$--$M$--$Y$ path [Fig.~\ref{Figure2}(b)]. The momentum-resolved spin splitting exhibits the characteristic $d$-wave pattern, $\Delta E_s(\mathbf{k})\propto k_x^2-k_y^2$, with opposite signs along the $k_x$ and $k_y$ directions in Fig.~\ref{Figure2}(c). The electron localization function (ELF) further identifies two spatially separated IAEs distributions within the lattice cavities [Fig.~\ref{Figure2}(d)], confirming the interstitial character of the electronic states. The spin-resolved projected density of states (PDOS) shows substantial contributions from both IAEs and neighboring Zr-$d$ orbitals near the Fermi level in Figs.~\ref{Figure2}(e) and (f), highlighting the significant contribution of IAEs to the low-energy electronic states ~\cite{Cheng2026InterstitialAM}.

An important consequence of nonrelativistic spin splitting is the generation of spin-polarized transport currents~\cite{GonzalezHernandez2021,Bai2022,Karube2022,Bai2023,Guo2026HybridParity}.
To establish the transport response of Zr$_2$N,
we calculate the spin conductivity within the semiclassical Boltzmann--Drude formalism under the constant relaxation-time approximation~\cite{PhysRevB.104.035408-Lz,MADSEN2018140,Ma2021}:
\begin{equation}
\sigma_{ij}^{\eta}
=
e^2\tau
\sum_n
\int_{\mathrm{BZ}}
\frac{d^2\mathbf{k}}{(2\pi)^2}
v_{n\mathbf{k},i}^{\eta}
v_{n\mathbf{k},j}^{\eta}
\left(
-\frac{\partial f}
{\partial\varepsilon_{n\mathbf{k}}^{\eta}}
\right),
\label{eq:drude_spin}
\end{equation}
where \(n\) is the band index, $\eta=\uparrow,\downarrow$ denotes the spin channel, $v_{n\mathbf{k},i}^{\eta}$ is the band group velocity, \(f\) is the Fermi–Dirac distribution function, $\varepsilon_{n\mathbf{k}}^{\eta}$ is the energy of the electronic state,  and $\tau$ is the carrier relaxation time, set to 10 fs as suggested by the reported relaxation time of electride Ca$_2$N~\cite{PhysRevB.98.155443}. 
The charge and spin conductivities are defined
as $\sigma_{ij}^{c}=\sigma_{ij}^{\uparrow}
+\sigma_{ij}^{\downarrow}$ and
$\sigma_{ij}^{s}=\sigma_{ij}^{\uparrow}
-\sigma_{ij}^{\downarrow}$, respectively.

\begin{figure}
\centering
\includegraphics[width=1\linewidth]{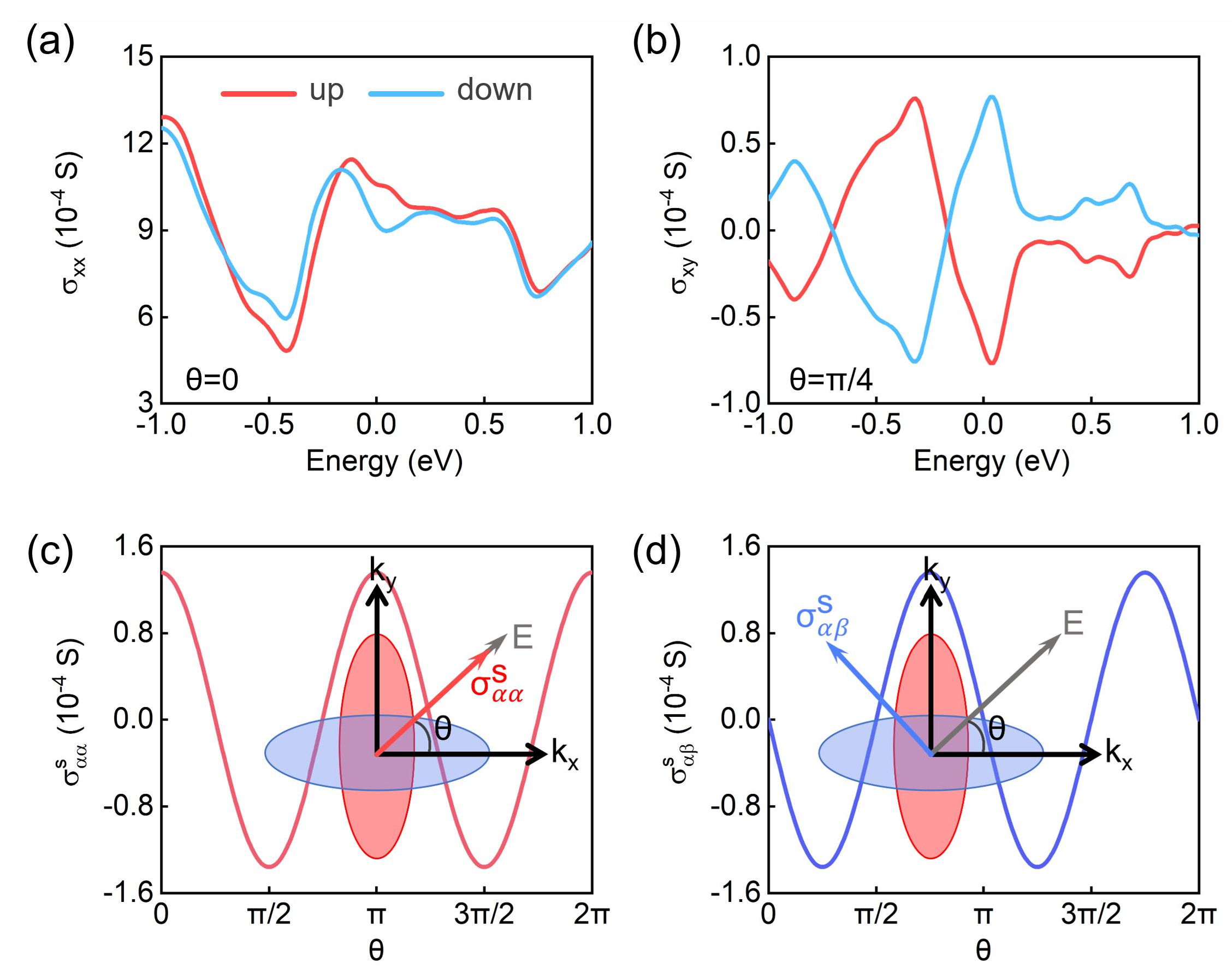}
\caption{\textbf{Spin-dependent transport in monolayer Zr$_2$N.}
(a, b) Spin-resolved longitudinal ($\sigma_{xx}$) and transverse ($\sigma_{xy}$) conductivities as functions of energy at $\theta=0$ and $\pi/4$, respectively.
(c, d) Angular dependence of the longitudinal ($\sigma_{\alpha\alpha}^{s}$) and transverse ($\sigma_{\alpha\beta}^{s}$) spin conductivities at the Fermi level, respectively. Here, $\theta$ denotes the angle between the applied electric field and the crystalline $x$ axis.
\label{Figure3}}
\end{figure}

The transport properties of Zr$_2$N are shown in Fig.~\ref{Figure3}. For an in-plane driving electric field
along the $x$ direction ($\theta=0$), the two spin channels
exhibit different longitudinal conductivities,
$\sigma_{xx}^{\uparrow}\neq\sigma_{xx}^{\downarrow}$,
as shown in Fig.~\ref{Figure3}(a). When the electric field is
oriented at $\theta=\pi/4$, the transverse conductivities
of opposite spins have equal magnitudes and opposite signs,
$\sigma_{xy}^{\uparrow}=-\sigma_{xy}^{\downarrow}$, as shown in Fig.~\ref{Figure3}(b). Consequently, the transverse charge current vanishes while a finite transverse spin current remains. 
This behavior follows directly from the $[C_2||S_4]$ symmetry which interchanges the $x$ and $y$ directions between opposite
spin channels, requiring $\sigma_{xx}^{\uparrow}=\sigma_{yy}^{\downarrow}$
and $\sigma_{yy}^{\uparrow}=\sigma_{xx}^{\downarrow}$.
These symmetry relations constrain the in-plane conductivity anisotropy to have opposite signs for the two spin channels, thereby explaining their different longitudinal responses along the same direction and oppositely signed transverse responses at $\theta=\pi/4$.
The resulting longitudinal and transverse spin conductivities exhibit characteristic
$\cos 2\theta$ and $\sin 2\theta$ angular dependences~\cite{Ma2021},
respectively, reflecting anisotropic spin transport, as shown in Figs.~\ref{Figure3}(c) and (d).

\begin{figure}
\centering
\includegraphics[width=1\linewidth]{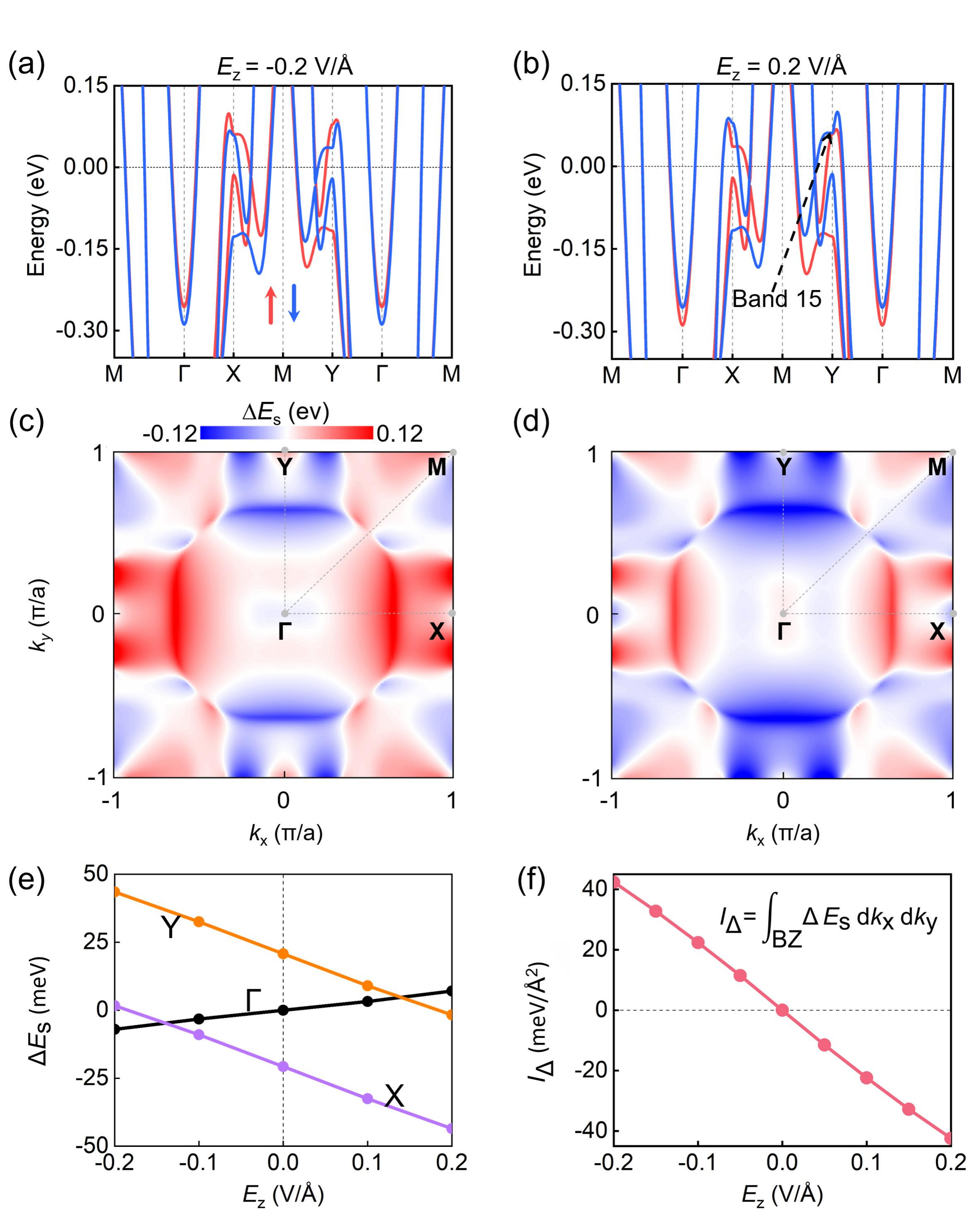}
\caption{\textbf{Electric field control of spin splitting in Zr$_2$N.}
(a, b) Spin-resolved band structures under out-of-plane electric fields of $E_z=-0.2$ and $0.2$~V/\AA, respectively.
(c, d) Corresponding momentum-resolved spin splitting, $\Delta E_s(\mathbf{k})=E_{\uparrow}(\mathbf{k})-E_{\downarrow}(\mathbf{k})$, for the selected band 15.
(e) Electric-field dependence of the spin splitting at the high-symmetry points $\Gamma$, X, and Y.
(f) Brillouin zone integrated spin splitting, $I_{\Delta}$, as a function of $E_z$.
\label{Figure4}}
\end{figure}

\subsection{Electric field induced magnetoelectric response}

The application of an out-of-plane electric field breaks the $[C_2||S_4]$ symmetry that connects the two spin-opposite IAEs sublattices, resulting in the two IAEs cavities energetically inequivalent. The resulting uncompensated ferrimagnetic configuration is no longer constrained to exhibit the pure $d$-wave spin splitting of the zero-field altermagnetic state. Importantly, although $[C_2||S_4]$ is broken at a fixed finite field, the corresponding
operation exchanges the two cavities and spin channels between opposite electric field, thereby relating their electronic and magnetic configurations.
This relation is reflected in the above TB model, when reversing the electric field interchanges the electron occupations and reverses the corresponding IAEs magnetic moments between the two cavities [Figs.~\ref{Figure1}(e) and (f)].

As shown in Fig.~\ref{Figure4}, these symmetry relations are reflected in the DFT calculated electronic structures. At $E_z=\pm0.2$~V/\AA, the spin-resolved band structures near the $X$ and $Y$ points are interchanged, accompanied by an exchange of the spin-up and spin-down channels upon reversing the electric field [Figs.~\ref{Figure4}(a) and (b)]. Moreover, the spin splitting gradually increases with increasing electric field [Fig. S1]. The momentum-resolved spin splitting in Figs.~\ref{Figure4}(c) and (d) further shows that the characteristic anisotropy persists, while the zero-field $d$-wave spin-splitting pattern is disrupted, driving a transition from altermagnetism to ferrimagnetism.
Importantly, reversing the electric field does not
simply reverse the spin splitting at the same momentum.
Instead, the $[C_2||S_4]$ operation
relates the spin-splitting distributions under opposite
electric fields through a simultaneous exchange of the
spin channels and a 90-degree rotation in momentum space,
requiring
$\Delta E_s(k_x,k_y,E_z)
=-\Delta E_s(-k_y,k_x,-E_z)$.

These features are quantitatively confirmed by the field dependence of the spin splitting, $\Delta E_s(\mathbf{k})=E_{\uparrow}(\mathbf{k})-E_{\downarrow}(\mathbf{k})$, at the high-symmetry points $\Gamma$, $X$, and $Y$ [Fig.~\ref{Figure4}(e)]. The field-induced splitting at $\Gamma$ reverses sign with $E_z$, while those at $X$ and $Y$ satisfy $\Delta E_{s{X}}(E_z)=-\Delta E_{s{Y}}(-E_z)$, consistent with the symmetry relation between opposite-field configurations. The Brillouin zone integrated spin splitting $I_{\Delta} = \int_{\text{BZ}} \Delta E_s \, dk_x \, dk_y$ exhibits an approximately linear dependence on $E_z$ and reverses sign upon field reversal in Fig.~\ref{Figure4}(f). Out-of-plane electric field thus represents a tuning knob for modulating the nonrelativistic spin splitting and spin transport of monolayer altermagnetic electrides.

\begin{figure}
\centering
\includegraphics[width=1\linewidth]{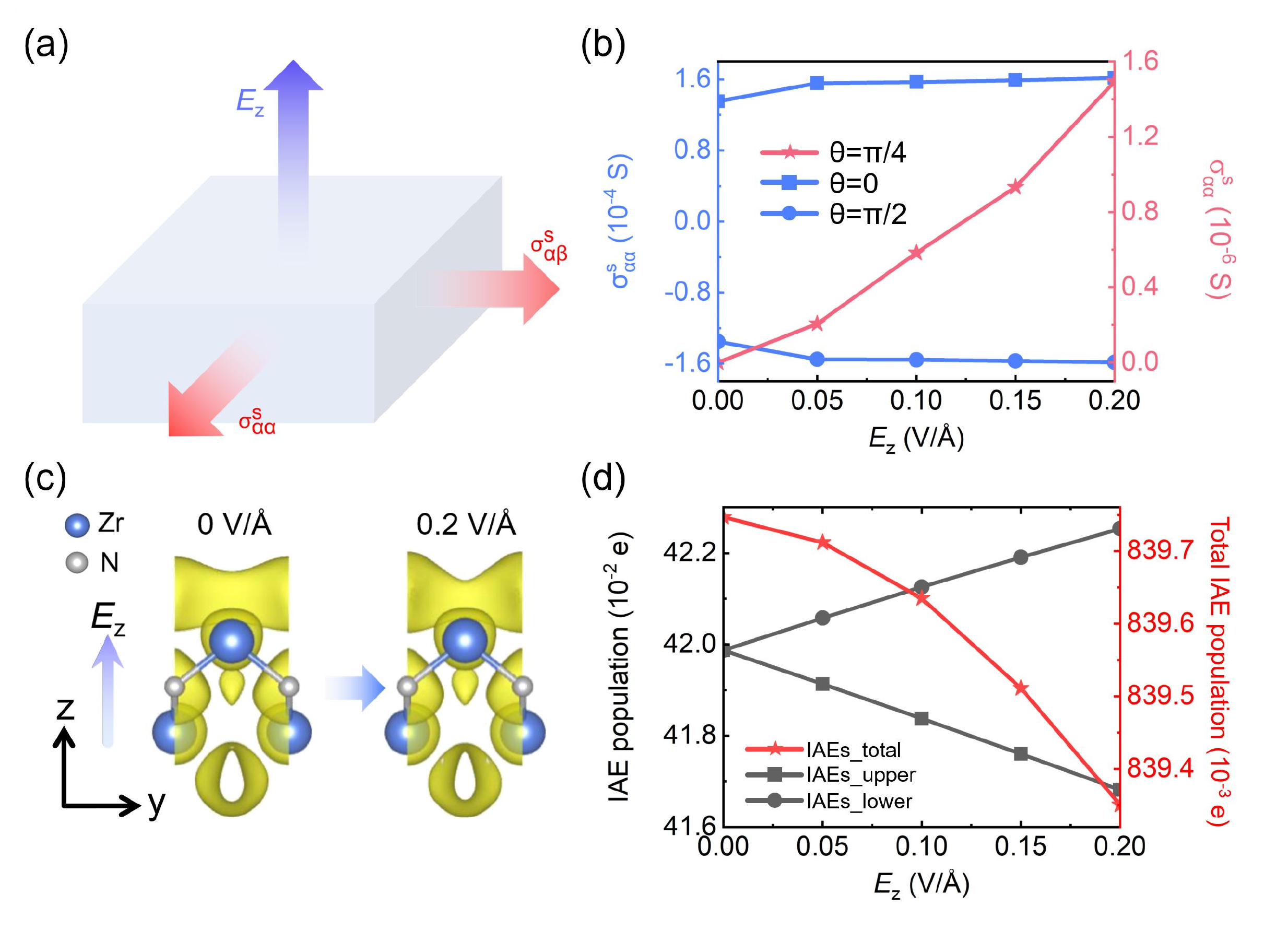}
\caption{\textbf{Electric field induced spin-polarized transport and electron redistribution.} 
(a) Schematic of the spin conductivity in the presence of
$E_z$. 
(b) The longitudinal spin conductivity at $\theta = 0, \pi/4$, and $\pi/2$ under the different $E_z$.
(c) The ELF of Zr$_2$N at $E_z=0$ and 0.2 V/\AA. The isosurface values are set as 0.3. 
(d) Electric field induced charge redistribution between the upper and lower IAEs for the different $E_z$.
\label{Figure5}}
\end{figure}

To further investigate the transport properties of the electric field induced spin splitting, we calculate the spin conductivity in the presence of $E_z$. We focus on both the longitudinal and transverse spin conductivity and their dependence on the electric field and transport direction [Fig.~\ref{Figure5}(a)]. Fig.~\ref{Figure5}(b) shows the longitudinal spin conductivity $\sigma^s_{\alpha\alpha}$ for the different $E_z$. As the electric field increases, $\sigma^s_{\alpha\alpha}$ at $\theta=\pi/4$ becomes nonzero, while the conductivities at $\theta=0$ and $\theta=\pi/2$ retain opposite signs but no longer have equal magnitudes. Moreover, the longitudinal and transverse spin conductivities retain their characteristic $\cos 2\theta$ and $\sin 2\theta$ angular dependences at $E_z = \pm0.2$ V/\AA, as shown in Fig.~S2. These behaviors arise from the electric field induced breaking of magnetic compensation between the two IAEs, which generates a finite net spin polarization and modulates the nonrelativistic spin splitting. These results further demonstrate that the out-of-plane electric field breaks the symmetry between the two interstitial cavities and disrupts the $d$-wave symmetry.

The microscopic origin of the magnetoelectric response induced by the electric field and the modulation of spin transport is analyzed through electron redistribution between the two cavities in Zr$_2$N, as shown in Figs.~\ref{Figure5}(c) and (d). The electric field induces an energy asymmetry between the two IAEs with opposite spins, driving electron redistribution and breaking magnetic compensation. The ELF exhibits symmetric electron localization in the two interstitial cavities at $E_z=0$~V/\AA, whereas an upward electric field of $0.2$ V/\AA~induces an asymmetric localization pattern [Fig.~\ref{Figure5}(c)], consistent with the TB model results. 

To quantitatively examine the electron redistribution indicated by the ELF results, the electron populations in the two interstitial cavities were calculated as
$N_{\alpha} = \int_{\Omega_{\alpha}} \rho(\mathbf{r})\,d^3r$. Here,  $\alpha=\mathrm{upper},\mathrm{lower}$, $\rho(\mathbf{r})$ is the electron density, and $\Omega_{\alpha}$ denotes the spherical integration region with a radius of 1.2~\AA~centered at the local ELF maxima of the corresponding cavity.
We find that an increasing upward electric field induces electron redistribution between the two cavities, decreasing the electron population in the upper IAEs while increasing that in the lower IAEs [Fig.~\ref{Figure5}(d)]. Moreover, their total electron population decreases only slightly.

\section{Conclusion}

In summary, we investigate interstitial electron mediated magnetoelectric response and its related spin-polarized transport properties in the altermagnetic electride Zr$_2$N. Based on the tight-binding model and first-principles calculations, we reveal that an out-of-plane electric field breaks the $d$-wave spin splitting symmetry and induces interstitial electron redistribution. The resulting magnetic imbalance drives a transition from altermagnetism to ferrimagnetism, generating a finite net spin polarization. The magnetoelectric response is further manifested in electrically tunable spin-polarized transport, providing complementary routes for manipulating spin-polarized carriers. Our findings reveal an interstitial electron mediated magnetoelectric effect in altermagnetic electrides. These results establish altermagnetic electrides as a viable building block for electrically controlled spintronic device technology.

\section{Computational Methods}

First-principles calculations were performed within the framework of density functional theory (DFT), as implemented in the Vienna Ab initio Simulation Package (VASP)~\cite{PhysRevB.54.11169,KRESSE199615}. The projector-augmented-wave (PAW) method was employed to describe the electron--ion interactions, and the Perdew--Burke--Ernzerhof (PBE) functional within the generalized gradient approximation (GGA) was used to describe the exchange--correlation interactions~\cite{PhysRevB.50.17953,PhysRevLett.77.3865}. The plane-wave energy cutoff was set to 500 eV. 
10$\times$10$\times$1 and 20$\times$20$\times$1 $\Gamma$-centered $k$-point meshes were adopted for the structural relaxation and electronic self-consistent calculations, respectively. 
The crystal structures of Zr$_2$N were fully relaxed for structural relaxation until the atomic forces were smaller than 0.01 eV/\AA~on each atomic site. The electronic iteration was performed until the total energy change was smaller than $10^{-6}$ eV.
A vacuum layer of 28 \AA~was introduced along the out-of-plane direction to avoid interactions between periodic images. $U_{\text{DFT}}=0$ eV was adopted in DFT calculation, and the calculated band structure is in good agreement with that previously reported in Ref.~\cite{Cheng2026InterstitialAM}.
The spin-dependent transport properties were calculated using the BoltzTraP2 package within the semiclassical Boltzmann transport framework and the constant relaxation-time approximation~\cite{MADSEN2018140}. A dense {130$\times$130$\times$1} $k$-point mesh was employed for electronic non-self-consistent calculations for the transport calculations, with a temperature of 300 K. Further details on the convergence tests are contained in the Supporting Information.

\begin{acknowledgments}
This work was supported by the Singapore Ministry of Education (MOE) Academic Research Fund (AcRF) Tier 2 grant (MOE-T2EP50125-0019). Y.S.A. acknowledges support from the Kwan Im Thong Hood Cho Temple Early Career Chair Professorship. J.L. acknowledges support from the National Natural Science Foundation of China (Grant No. 12674285) and the National Key R\&D Program of China (Grant No. 2022YFA1203904). Y.D. acknowledges support from the National Natural Science Foundation of China (Grant No. 12274016) and the Fundamental Research Funds for the Central Universities (Grant Nos. 501XSKC2025119001 and YWF-22-K-101). J.F.W. acknowledges support from the National Natural Science Foundation of China (Grant No. 12474217), the Beijing Natural Science Foundation (Grant No. 1242023), and the Fundamental Research Funds for the Central Universities (Grant No. 501XYGG2026119003). W.M. acknowledges support from the Science Research Project of Hebei Education Department (Grant No. BJ2025051). Q.L. acknowledges support from the Science and Technology Innovation Project for Youth Talents of Enshi (Grant No. D20220066) and the High-Level Research Achievement Cultivation Project of Hubei Minzu University (Grant No. XN2321). Y.L., Q.L., and J.H.W. acknowledge financial support from the China Scholarship Council (CSC). We acknowledge the High-Performance Computing Platform of Peking University for providing computational resources.

\end{acknowledgments}

\section*{Conflicts of Interest}

The authors declare no conflicts of interest.

\section*{Data Availability Statement}

The data that support the findings of this study are available
from the corresponding author upon reasonable request.

\bibliographystyle{apsrev4-2}
\bibliography{ref_AS_full_journals_corrected}

@article{peng2025ferroelastic,
  author  = {Peng, Rui and Fang, Shibo and Ho, Pin and Liu, Fanxin and Zhou, Tong and Liu, Junwei and Ang, Yee Sin},
  title   = {Ferroelastic Altermagnetism},
  journal = {npj Quantum Materials},
  volume  = {11},
  number  = {1},
  pages   = {5},
  year    = {2026},
  unused_doi     = {10.1038/s41535-025-00835-7}
}

@article{10.1021/jacs.5c16402,
    author = {Che, Yixuan and Guo, Yuhang and Lv, Haifeng and Wu, Xiaojun and Yang, Jinlong},
    title = {Symmetry-Driven Multiferroic Altermagnetism in Two-Dimensional Materials},
    journal = {Journal of the American Chemical Society},
    volume = {148},
    number = {5},
    pages = {5125},
    year = {2026},
    unused_doi = {10.1021/jacs.5c16402},
    unused_url = {https://doi.org/10.1021/jacs.5c16402}
}

@article{PRL2026,
  title = {Origin of Electride States and Their Responses under Pressure},
  author = {Cai, Xuefen and Huang, Qiu-Shi and Wei, Su-Huai},
  journal = {Phys. Rev. Lett.},
  volume = {137},
  issue = {21},
  pages = {216402},
  numpages = {6},
  year = {2026},
  month = {Nov},
  publisher = {American Physical Society},
  unused_doi = {10.1103/63ch-m64p},
  unused_url = {https://link.aps.org/doi/10.1103/63ch-m64p}
}

@article{Lee2013Ca2N,
  author = {Lee, Kimoon and Kim, Sung Wng and Toda, Yoshitake and Matsuishi, Satoru and Hosono, Hideo},
  title = {Dicalcium Nitride as a Two-Dimensional Electride with an Anionic Electron Layer},
  journal = {Nature},
  volume = {494},
  pages = {336--340},
  year = {2013},
  unused_doi = {10.1038/nature11812}
}

@article{HosonoKitano2021Electrides,
  author = {Hosono, Hideo and Kitano, Masaaki},
  title = {Advances in Materials and Applications of Inorganic Electrides},
  journal = {Chemical Reviews},
  volume = {121},
  pages = {3121--3185},
  year = {2021},
  unused_doi = {10.1021/acs.chemrev.0c01071}
}

@article{Lim2024,
  author  = {Lim, Chan-Young and Kim, Min-Seok and Lim, Dong Cheol and Kim, Sunghun and Lee, Yeonghoon and Cha, Jaehoon and Lee, Gyubin and Song, Sang Yong and Thapa, Dinesh and Denlinger, Jonathan D. and Kim, Seong-Gon and Kim, Sung Wng and Seo, Jungpil and Kim, Yeongkwan},
  title   = {Topological Fermi-Arc Surface State Covered by Floating Electrons on a Two-Dimensional Electride},
  journal = {Nature Communications},
  volume  = {15},
  pages   = {5615},
  year    = {2024},
  unused_doi     = {10.1038/s41467-024-49841-6}
}

@article{6dv1-7ddd,
  title = {Obstructed Atomic Insulator Inorganic Electrides},
  author = {Li, Yang and Wang, Jianhua and Qian, Shifeng and Meng, Weizhen and Wang, Xiaotian},
  journal = {Physical Review B},
  volume = {112},
  issue = {20},
  pages = {205112},
  numpages = {11},
  year = {2025},
  month = {Nov},
  publisher = {American Physical Society},
  unused_doi = {10.1103/6dv1-7ddd},
  unused_url = {https://link.aps.org/doi/10.1103/6dv1-7ddd}
}

@article{Geng2026,
  author  = {Geng, Songyuan and Wang, Xin and Guo, Risi and Qiu, Chen and Chen, Fangjie and Wang, Qun and Li, Kangjie and Hao, Peipei and Liang, Hanpu and Huang, Yang and Wu, Yunbo and Cui, Shengtao and Sun, Zhe and Kim, Timur K. and Cacho, Cephise and Dessau, Daniel S. and Zhou, Benjamin T. and Li, Haoxiang},
  title   = {Experimental Realization of Dice-Lattice Flat Band at the Fermi Level in Layered Electride {YCl}},
  journal = {Nature Communications},
  volume  = {17},
  pages   = {2213},
  year    = {2026},
  unused_doi     = {10.1038/s41467-026-69049-0}
}

@article{Miyakawa2007ElectrideSC,
  author = {Miyakawa, Masashi and Kim, Sung Wng and Hirano, Masahiro and Kohama, Yoshimitsu and Kawaji, Hitoshi and Atake, Tooru and Ikegami, Hiroki and Kono, Kimitoshi and Hosono, Hideo},
  title = {Superconductivity in an Inorganic Electride {12CaO$\cdot$7Al$_2$O$_3$:e$^-$}},
  journal = {Journal of the American Chemical Society},
  volume = {129},
  pages = {7270--7271},
  year = {2007},
  unused_doi = {10.1021/ja0724644}
}

@article{Zhang2017,
  author  = {Zhang, Yaoqing and Wang, Bosen and Xiao, Zewen and Lu, Yangfan and Kamiya, Toshio and Uwatoko, Yoshiya and Kageyama, Hiroshi and Hosono, Hideo},
  title   = {Electride and Superconductivity Behaviors in {Mn5Si3}-Type Intermetallics},
  journal = {npj Quantum Materials},
  year    = {2017},
  volume  = {2},
  number  = {1},
  pages   = {45},
  unused_doi     = {10.1038/s41535-017-0053-4},
  unused_url     = {https://doi.org/10.1038/s41535-017-0053-4}
}

@article{Ma2009DenseSodium,
  author = {Ma, Yanming and Eremets, Mikhail and Oganov, Artem R. and Xie, Yu and Trojan, Ivan and Medvedev, Sergey and Lyakhov, Andriy O. and Valle, Mario and Prakapenka, Vitali},
  title = {Transparent Dense Sodium},
  journal = {Nature},
  volume = {458},
  pages = {182--185},
  year = {2009},
  unused_doi = {10.1038/nature07786}
}

@article{mengAM2025,
author = {Meng, Weizhen and Tian, Lu and Zhou, Feng and Mo, Zhaojun and Jiao, Yalong and Wang, Shiyao and Jiang, Jiayu and Zhang, Xiaoming and Cheng, Zhenxiang and Liu, Ying and Wang, Wenhong and Zhang, Gang and Wang, Xiaotian},
title = {1D Magnetic Topological Inorganic Electrides},
journal = {Advanced Materials},
volume = {37},
number = {19},
pages = {2418904},
year = {2025},
unused_doi = {https://doi.org/10.1002/adma.202418904}
}

@article{PhysRevLett.125.187203,
  title = {Ferromagnetic Weyl Fermions in Two-Dimensional Layered Electride ${\mathrm{Gd}}_{2}\mathrm{C}$},
  author = {Liu, Shuyuan and Wang, Chongze and Liu, Liangliang and Choi, Jin-Ho and Kim, Hyun-Jung and Jia, Yu and Park, Chul Hong and Cho, Jun-Hyung},
  journal = {Physical Review Letters},
  volume = {125},
  issue = {18},
  pages = {187203},
  numpages = {6},
  year = {2020},
  month = {Oct},
  publisher = {American Physical Society},
  unused_doi = {10.1103/PhysRevLett.125.187203},
  unused_url = {https://link.aps.org/doi/10.1103/PhysRevLett.125.187203}
}

@article{Kitano2012Ammonia,
  author = {Kitano, Masaaki and Inoue, Yasunori and Yamazaki, Youhei and Hayashi, Fumitaka and Kanbara, Shinji and Matsuishi, Satoru and Yokoyama, Toshiharu and Kim, Sung-Wng and Hara, Michikazu and Hosono, Hideo},
  title = {Ammonia Synthesis Using a Stable Electride as an Electron Donor and Reversible Hydrogen Store},
  journal = {Nature Chemistry},
  volume = {4},
  pages = {934--940},
  year = {2012},
  unused_doi = {10.1038/nchem.1476}
}

@article{Kitano2015,
  author  = {Kitano, Masaaki and Kanbara, Shinji and Inoue, Yasunori and Kuganathan, Navaratnarajah and Sushko, Peter V. and Yokoyama, Toshiharu and Hara, Michikazu and Hosono, Hideo},
  title   = {Electride Support Boosts Nitrogen Dissociation over Ruthenium Catalyst and Shifts the Bottleneck in Ammonia Synthesis},
  journal = {Nature Communications},
  volume  = {6},
  pages   = {6731},
  year    = {2015},
  unused_doi     = {10.1038/ncomms7731}
}

@article{Guan2017NegativeRefraction,
  author = {Guan, Shan and Huang, Shao Ying and Yao, Yugui and Yang, Shengyuan A.},
  title = {Tunable Hyperbolic Dispersion and Negative Refraction in Natural Electride Materials},
  journal = {Physical Review B},
  volume = {95},
  pages = {165436},
  year = {2017},
  unused_doi = {10.1103/PhysRevB.95.165436}
}

@article{Burton2018ElectrideScreening,
  author = {Burton, Lee A. and Ricci, Francesco and Chen, Wei and Rignanese, Gian-Marco and Hautier, Geoffroy},
  title = {High-Throughput Identification of Electrides from All Known Inorganic Materials},
  journal = {Chemistry of Materials},
  volume = {30},
  pages = {7521--7526},
  year = {2018},
  unused_doi = {10.1021/acs.chemmater.8b02526}
}

@article{Meng2024ElectrideReview,
  author = {Meng, Weizhen and Wang, Jianhua and Wang, Xiaotian and Wang, Wenhong and Zhang, Xiaoming and Bando, Yoshio and Cheng, Zhenxiang},
  title = {Multi-Dimensional Inorganic Electrides for Energy Conversion and Storage},
  journal = {Journal of Materials Chemistry A},
  volume = {12},
  pages = {2583--2604},
  year = {2024},
  unused_doi = {10.1039/D3TA06546D}
}

@article{Zhang2023MagneticElectrides,
  author = {Zhang, Xiaoming and Meng, Weizhen and Liu, Ying and Dai, Xuefang and Liu, Guodong and Kou, Liangzhi},
  title = {Magnetic Electrides: High-Throughput Material Screening, Intriguing Properties, and Applications},
  journal = {Journal of the American Chemical Society},
  volume = {145},
  pages = {5523--5535},
  year = {2023},
  unused_doi = {10.1021/jacs.3c00284}
}

@article{Meng2025MagneticElectrideReview,
  author = {Meng, Weizhen and Bai, Jingbo and Ma, Fengxian and Jiao, Yalong and Wang, Shiyao and Jiang, Jiayu and Zhang, Xiaoming and Cheng, Zhenxiang and Yang, Tie},
  title = {Magnetic Electrides: Recent Advances in Materials Realization and Application Prospects},
  journal = {Applied Physics Reviews},
  volume = {12},
  pages = {011320},
  year = {2025},
  unused_doi = {10.1063/5.0233622}
}

@article{Smejkal2022Beyond,
  author = {{\v{S}}mejkal, Libor and Sinova, Jairo and Jungwirth, Tomas},
  title = {Beyond Conventional Ferromagnetism and Antiferromagnetism: A Phase with Nonrelativistic Spin and Crystal Rotation Symmetry},
  journal = {Physical Review X},
  volume = {12},
  pages = {031042},
  year = {2022},
  unused_doi = {10.1103/PhysRevX.12.031042}
}

@article{Smejkal2022Landscape,
  author = {{\v{S}}mejkal, Libor and Sinova, Jairo and Jungwirth, Tomas},
  title = {Emerging Research Landscape of Altermagnetism},
  journal = {Physical Review X},
  volume = {12},
  pages = {040501},
  year = {2022},
  unused_doi = {10.1103/PhysRevX.12.040501}
}

@article{Krempasky2024Altermagnetic,
  author = {Krempask{\'y}, J. and {\v{S}}mejkal, L. and D'Souza, S. W. and Hajlaoui, M. and Springholz, G. and Uhl{\'i}{\v{r}}ov{\'a}, K. and Alarab, F. and Constantinou, P. C. and Strocov, V. and Usanov, D. and Pudelko, W. R. and Gonz{\'a}lez-Hern{\'a}ndez, R. and Hellenes, A. B. and Jansa, Z. and Reichlov{\'a}, H. and {\v{S}}ob{\'a}{\v{n}}, Z. and Gonzalez Betancourt, R. D. and Wadley, P. and Sinova, J. and Kriegner, D. and Min{\'a}r, J. and Dil, J. H. and Jungwirth, T.},
  title = {Altermagnetic Lifting of Kramers Spin Degeneracy},
  journal = {Nature},
  volume = {626},
  pages = {517--522},
  year = {2024},
  unused_doi = {10.1038/s41586-023-06907-7}
}

@article{PhysRevLett.133.206401-2024,
  title = {Large Band Splitting in $g$-Wave Altermagnet CrSb},
  author = {Ding, Jianyang and Jiang, Zhicheng and Chen, Xiuhua and Tao, Zicheng and Liu, Zhengtai and Li, Tongrui and Liu, Jishan and Sun, Jianping and Cheng, Jinguang and Liu, Jiayu and Yang, Yichen and Zhang, Runfeng and Deng, Liwei and Jing, Wenchuan and Huang, Yu and Shi, Yuming and Ye, Mao and Qiao, Shan and Wang, Yilin and Guo, Yanfeng and Feng, Donglai and Shen, Dawei},
  journal = {Physical Review Letters},
  volume = {133},
  issue = {20},
  pages = {206401},
  numpages = {7},
  year = {2024},
  month = {Nov},
  publisher = {American Physical Society},
  unused_doi = {10.1103/PhysRevLett.133.206401},
  unused_url = {https://link.aps.org/doi/10.1103/PhysRevLett.133.206401}
}

@article{PhysRevLett.132.036702-2024,
  title = {Broken Kramers Degeneracy in Altermagnetic MnTe},
  author = {Lee, Suyoung and Lee, Sangjae and Jung, Saegyeol and Jung, Jiwon and Kim, Donghan and Lee, Yeonjae and Seok, Byeongjun and Kim, Jaeyoung and Park, Byeong Gyu and {\ifmmode \check{S}\else \v{S}\fi{}}mejkal, Libor and Kang, Chang-Jong and Kim, Changyoung},
  journal = {Physical Review Letters},
  volume = {132},
  issue = {3},
  pages = {036702},
  numpages = {7},
  year = {2024},
  month = {Jan},
  publisher = {American Physical Society},
  unused_doi = {10.1103/PhysRevLett.132.036702},
  unused_url = {https://link.aps.org/doi/10.1103/PhysRevLett.132.036702}
}

@article{Jungwirth2026Altermagnetism,
  author = {Jungwirth, Tomas and Sinova, Jairo and Fernandes, Rafael M. and Liu, Qihang and Watanabe, Hikaru and Murakami, Shuichi and Nakatsuji, Satoru and {\v{S}}mejkal, Libor},
  title = {Symmetry, Microscopy and Spectroscopy Signatures of Altermagnetism},
  journal = {Nature},
  volume = {649},
  pages = {837--847},
  year = {2026},
  unused_doi = {10.1038/s41586-025-09883-2}
}

@article{Yu2026,
  author  = {Yu, Cuiju and Lado, Jose L.},
  title   = {Intrinsic {$i$}-Wave Altermagnetism in 2D Graphene Superlattices},
  journal = {Advanced Science},
  pages   = {e78167},
  year    = {2026},
  unused_doi     = {10.1002/advs.78167}
}

@article{PhysRevLett.134.106801,
  title = {Antiferroelectric Altermagnets: Antiferroelectricity Alters Magnets},
  author = {Duan, Xunkai and Zhang, Jiayong and Zhu, Ziye and Liu, Yuntian and Zhang, Zhenyu and {\ifmmode \check{Z}\else \v{Z}\fi{}}uti{\ifmmode \acute{c}\else \'{c}\fi{}}, Igor and Zhou, Tong},
  journal = {Physical Review Letters},
  volume = {134},
  issue = {10},
  pages = {106801},
  numpages = {8},
  year = {2025},
  month = {Mar},
  publisher = {American Physical Society},
  unused_doi = {10.1103/PhysRevLett.134.106801},
  unused_url = {https://link.aps.org/doi/10.1103/PhysRevLett.134.106801}
}

@article{PhysRevLett.134.106802-liu,
  title = {Ferroelectric Switchable Altermagnetism},
  author = {Gu, Mingqiang and Liu, Yuntian and Zhu, Haiyuan and Yananose, Kunihiro and Chen, Xiaobing and Hu, Yongkang and Stroppa, Alessandro and Liu, Qihang},
  journal = {Physical Review Letters},
  volume = {134},
  issue = {10},
  pages = {106802},
  numpages = {7},
  year = {2025},
  month = {Mar},
  publisher = {American Physical Society},
  unused_doi = {10.1103/PhysRevLett.134.106802},
  unused_url = {https://link.aps.org/doi/10.1103/PhysRevLett.134.106802}
}

@article{Zhu2024,
  author  = {Zhu, Yu-Peng and Chen, Xiaobing and Liu, Xiang-Rui and Liu, Yuntian and Liu, Pengfei and Zha, Heming and Qu, Gexing and Hong, Caiyun and Li, Jiayu and Jiang, Zhicheng and Ma, Xiao-Ming and Hao, Yu-Jie and Zhu, Ming-Yuan and Liu, Wenjing and Zeng, Meng and Jayaram, Sreehari and Lenger, Malik and Ding, Jianyang and Mo, Shu and Tanaka, Kiyohisa and Arita, Masashi and Liu, Zhengtai and Ye, Mao and Shen, Dawei and Wrachtrup, J{\"o}rg and Huang, Yaobo and He, Rui-Hua and Qiao, Shan and Liu, Qihang and Liu, Chang},
  title   = {Observation of Plaid-Like Spin Splitting in a Noncoplanar Antiferromagnet},
  journal = {Nature},
  volume  = {626},
  number  = {7999},
  pages   = {523--528},
  year    = {2024},
  unused_doi     = {10.1038/s41586-024-07023-w}
}

@article{Zhou2025,
  author  = {Zhou, Zhiyuan and Cheng, Xingkai and Hu, Mengli and Chu, Ruiyue and Bai, Hua and Han, Lei and Liu, Junwei and Pan, Feng and Song, Cheng},
  title   = {Manipulation of the Altermagnetic Order in {CrSb} via Crystal Symmetry},
  journal = {Nature},
  volume  = {638},
  number  = {8051},
  pages   = {645--650},
  year    = {2025},
  unused_doi     = {10.1038/s41586-024-08436-3}
}

@article{Wang2026Pentagonal,
  author = {Wang, Jianhua and Yang, Xingyue and Yang, Zongmeng and Lu, Jing and Ho, Pin and Wang, Wenhong and Ang, Yee Sin and Cheng, Zhenxiang and Fang, Shibo},
  title = {Pentagonal {2D} Altermagnets: Material Screening and Altermagnetic Tunneling Junction Device Application},
  journal = {Advanced Functional Materials},
  volume = {36},
  pages = {2505145},
  year = {2026},
  unused_doi = {10.1002/adfm.202505145}
}

@article{Fang2026EdgeStates,
  author = {Fang, Shibo and Yang, Zongmeng and Wang, Jianhua and Yang, Xingyue and Lu, Jing and Lee, Ching Hua and Wang, Xiaotian and Ang, Yee Sin},
  title = {One-Dimensional Electronics with Edge States in Two-Dimensional Altermagnets},
  journal = {Physical Review B},
  volume = {113},
  pages = {224407},
  year = {2026},
  unused_doi = {10.1103/d1gz-fmwy}
}

@article{Cheng2026InterstitialAM,
  author  = {Cheng, Xia and Wu, Yang and Guo, Zhenzhou and Yang, Tie and Meng, Weizhen and Cheng, Zhenxiang and Yu, Zhi-Ming and Wang, Xiaotian},
  title   = {Interstitial-Electron Altermagnetism in Two Dimensions},
  journal = {arXiv preprint arXiv:2603.26404},
  year    = {2026},
  unused_doi     = {10.48550/arXiv.2603.26404}
}

@article{Lottermoser2004,
  title   = {Magnetic Phase Control by an Electric Field},
  author  = {Lottermoser, Thomas and Lonkai, Thomas and Amann, Uwe and Hohlwein, Dietmar and Ihringer, J{\"o}rg and Fiebig, Manfred},
  journal = {Nature},
  volume  = {430},
  number  = {6999},
  pages   = {541--544},
  year    = {2004},
  unused_doi     = {10.1038/nature02728},
  unused_url     = {https://doi.org/10.1038/nature02728}
}

@article{Eerenstein2006,
  author  = {Eerenstein, W. and Mathur, N. D. and Scott, J. F.},
  title   = {Multiferroic and Magnetoelectric Materials},
  journal = {Nature},
  volume  = {442},
  number  = {7104},
  pages   = {759--765},
  year    = {2006},
  unused_doi     = {10.1038/nature05023}
}

@article{Eerenstein2007,
  author  = {Eerenstein, W. and Wiora, M. and Prieto, J. L. and Scott, J. F. and Mathur, N. D.},
  title   = {Giant Sharp and Persistent Converse Magnetoelectric Effects in Multiferroic Epitaxial Heterostructures},
  journal = {Nature Materials},
  volume  = {6},
  number  = {5},
  pages   = {348--351},
  year    = {2007},
  unused_doi     = {10.1038/nmat1886}
}

@article{PhysRevLett.104.127202,
  title = {Origin of the Magnetoelectric Coupling Effect in $\mathrm{Pb}({\mathrm{Zr}}_{0.2}{\mathrm{Ti}}_{0.8})\mathrm{O}_{3}/\mathrm{La}_{0.8}\mathrm{Sr}_{0.2}\mathrm{MnO}_{3}$ Multiferroic Heterostructures},
  author = {Vaz, C. A. F. and Hoffman, J. and Segal, Y. and Reiner, J. W. and Grober, R. D. and Zhang, Z. and Ahn, C. H. and Walker, F. J.},
  journal = {Physical Review Letters},
  volume = {104},
  issue = {12},
  pages = {127202},
  numpages = {4},
  year = {2010},
  month = {Mar},
  publisher = {American Physical Society},
  unused_doi = {10.1103/PhysRevLett.104.127202},
  unused_url = {https://link.aps.org/doi/10.1103/PhysRevLett.104.127202}
}

@article{Yang2014,
  author  = {Yang, Sheng-Wei and Peng, Ren-Ci and Jiang, Tao and Liu, Yu-Kuai and Feng, Lei and Wang, Jian-Jun and Chen, Long-Qing and Li, Xiao-Guang and Nan, Ce-Wen},
  title   = {Non-Volatile 180{$^\circ$} Magnetization Reversal by an Electric Field in Multiferroic Heterostructures},
  journal = {Advanced Materials},
  volume  = {26},
  number  = {41},
  pages   = {7091--7095},
  year    = {2014},
  unused_doi     = {10.1002/adma.201402774}
}

@article{Spaldin2019,
  author  = {Spaldin, N. A. and Ramesh, R.},
  title   = {Advances in Magnetoelectric Multiferroics},
  journal = {Nature Materials},
  volume  = {18},
  number  = {3},
  pages   = {203--212},
  year    = {2019},
  unused_doi     = {10.1038/s41563-018-0275-2}
}

@article{Zhang2024GateField,
  author = {Zhang, Run-Wu and Cui, Chaoxi and Li, Runze and Duan, Jingyi and Li, Lei and Yu, Zhi-Ming and Yao, Yugui},
  title = {Predictable Gate-Field Control of Spin in Altermagnets with Spin-Layer Coupling},
  journal = {Physical Review Letters},
  volume = {133},
  pages = {056401},
  year = {2024},
  unused_doi = {10.1103/PhysRevLett.133.056401}
}

@article{Peng2025LayerSpin,
  author = {Peng, Rui and Yang, Jin and Hu, Lin and Ong, Wee-Liat and Ho, Pin and Lau, Chit Siong and Liu, Junwei and Ang, Yee Sin},
  title = {All-Electrical Layer-Spintronics in Altermagnetic Bilayers},
  journal = {Materials Horizons},
  volume = {12},
  pages = {2197--2207},
  year = {2025},
  unused_doi = {10.1039/D4MH01509F}
}

@article{Lee2020MagneticElectride,
  author = {Lee, Seung Yong and Hwang, Jae-Yeol and Park, Jongho and Nandadasa, Chandani N. and Kim, Younghak and Bang, Joonho and Lee, Kimoon and Lee, Kyu Hyoung and Zhang, Yunwei and Ma, Yanming and Hosono, Hideo and Lee, Young Hee and Kim, Seong-Gon and Kim, Sung Wng},
  title = {Ferromagnetic Quasi-Atomic Electrons in Two-Dimensional Electride},
  journal = {Nature Communications},
  volume = {11},
  pages = {1526},
  year = {2020},
  unused_doi = {10.1038/s41467-020-15253-5}
}

@article{Lee2021CavityFerrimagnetism,
  author = {Lee, Seung Yong and Bang, Joonho and Song, Hyun Yong and Yoo, Byung Il and Kim, Yeongkwan and Lee, Kimoon and Thapa, Dinesh and Kim, Seong-Gon and Kim, Sung Wng},
  title = {Mixed-Cation Driven Magnetic Interaction of Interstitial Electrons for Ferrimagnetic Two-Dimensional Electride},
  journal = {npj Quantum Materials},
  volume = {6},
  pages = {21},
  year = {2021},
  unused_doi = {10.1038/s41535-021-00319-4}
}

@article{Lee2023ElectrideHydride,
  author = {Lee, Seung Yong and Lim, Dong Cheol and Khan, Md Salman and Hwang, Jeong Yun and Kim, Hyung Sub and Lee, Kyu Hyung and Kim, Sung Wng},
  title = {Magnetic Quasi-Atomic Electrons Driven Reversible Structural and Magnetic Transitions between Electride and Its Hydrides},
  journal = {Nature Communications},
  volume = {14},
  pages = {5469},
  year = {2023},
  unused_doi = {10.1038/s41467-023-41085-0}
}

@article{Hubbard1963,
  author  = {Hubbard, J.},
  title   = {Electron Correlations in Narrow Energy Bands},
  journal = {Proceedings of the Royal Society of London. Series A, Mathematical and Physical Sciences},
  volume  = {276},
  pages   = {238--257},
  year    = {1963},
  unused_doi     = {10.1098/rspa.1963.0204}
}

@article{Bach1994,
  author  = {Bach, Volker and Lieb, Elliott H. and Solovej, Jan Philip},
  title   = {Generalized Hartree-Fock Theory and the Hubbard Model},
  journal = {Journal of Statistical Physics},
  volume  = {76},
  number  = {1},
  pages   = {3--89},
  year    = {1994},
  unused_doi     = {10.1007/BF02188656}
}

@article{Zha2025CrossChainElectride,
  author  = {Zha, Da-Bao and Jiang, Peng and Li, Yan-Ling and Lin, Hai-Qing},
  title   = {Robust Cross-Chain Surface Interstitial Electronic States and Doping-Enhanced Superconductivity in Monolayer {$M_2$N} ({$M$ = Ti, Zr, Hf}) Electrides},
  journal = {Physical Review B},
  volume  = {112},
  pages   = {115416},
  year    = {2025},
  unused_doi     = {10.1103/s754-h6v9}
}

@article{Yu2026FINDSPINGROUP,
  title   = {Identifying Oriented Spin Space Groups and Related Physical Properties Using an Online Platform FINDSPINGROUP},
  author  = {Yu, Yutong and Chen, Xiaobing and Zhu, Yanzhou and Li, Yuhui and Xiong, Renzheng and Li, Jiayu and Liu, Yuntian and Liu, Qihang},
  journal = {National Science Review},
  year    = {2026},
  pages   = {nwag614},
  unused_url     = {https://doi.org/10.1093/nsr/nwag614}
}

@article{GonzalezHernandez2021,
  author  = {Gonz{\'a}lez-Hern{\'a}ndez, Rafael and {\v{S}}mejkal, Libor and V{\'y}born{\'y}, Karel and Yahagi, Yuta and Sinova, Jairo and Jungwirth, Tom{\'a}{\v{s}} and {\v{Z}}elezn{\'y}, Jakub},
  title   = {Efficient Electrical Spin Splitter Based on Nonrelativistic Collinear Antiferromagnetism},
  journal = {Physical Review Letters},
  volume  = {126},
  pages   = {127701},
  year    = {2021},
  unused_doi     = {10.1103/PhysRevLett.126.127701}
}

@article{Bai2022,
  author  = {Bai, Hua and Han, Lei and Feng, Xiaoyu and Zhou, Yijun and Su, Ruixia and Wang, Qian and Liao, Liyue and Zhu, Wenxuan and Chen, Xinzhi and Pan, Feng and Fan, Xiaolong and Song, Cheng},
  title   = {Observation of Spin Splitting Torque in a Collinear Antiferromagnet {RuO$_2$}},
  journal = {Physical Review Letters},
  volume  = {128},
  pages   = {197202},
  year    = {2022},
  unused_doi     = {10.1103/PhysRevLett.128.197202}
}

@article{Karube2022,
  author  = {Karube, Shutaro and Tanaka, Takahiro and Sugawara, Daichi and Kadoguchi, Naohiro and Kohda, Makoto and Nitta, Junsaku},
  title   = {Observation of Spin-Splitter Torque in Collinear Antiferromagnetic {RuO$_2$}},
  journal = {Physical Review Letters},
  volume  = {129},
  pages   = {137201},
  year    = {2022},
  unused_doi     = {10.1103/PhysRevLett.129.137201}
}

@article{Bai2023,
  author  = {Bai, Hua and Zhang, Y. C. and Zhou, Y. J. and Chen, P. and Wan, C. H. and Han, L. and Zhu, W. X. and Liang, S. X. and Su, Y. C. and Han, X. F. and Pan, F. and Song, C.},
  title   = {Efficient Spin-to-Charge Conversion via Altermagnetic Spin Splitting Effect in Antiferromagnet {RuO$_2$}},
  journal = {Physical Review Letters},
  volume  = {130},
  pages   = {216701},
  year    = {2023},
  unused_doi     = {10.1103/PhysRevLett.130.216701}
}

@article{Guo2026HybridParity,
    author = {Guo, Zhenzhou and Yu, Jiangtao and Fang, Shibo and Cao, Jin and Zhou, Xiaodong and Ang, Yee Sin and Wang, Wenhong and Cheng, Zhenxiang and Wang, Xiaotian},
    title = {Hybrid-Parity Sliding Multiferroics},
    journal = {Nature Communications},
    year = {2026},
    month = {11},
    unused_doi = {10.1038/s41467-026-79435-3},
   unused_url = {https://doi.org/10.1038/s41467-026-79435-3}
}

@article{PhysRevB.104.035408-Lz,
  title = {Fermiology of Two-Dimensional Titanium Carbide and Nitride MXenes},
  author = {Bagheri, Mohammad and Ibragimova, Rina and Komsa, Hannu-Pekka},
  journal = {Physical Review B},
  volume = {104},
  issue = {3},
  pages = {035408},
  numpages = {7},
  year = {2021},
  month = {Jul},
  publisher = {American Physical Society},
  unused_doi = {10.1103/PhysRevB.104.035408},
  unused_url = {https://link.aps.org/doi/10.1103/PhysRevB.104.035408}
}

@article{PhysRevB.98.155443,
  title = {Electron-Phonon Interaction in a ${\mathrm{Ca}}_{2}\mathrm{N}$ Monolayer: Intrinsic Mobility of Electrene},
  author = {Zeng, Xiongzhi and Zhao, Songtao and Li, Zhenyu and Yang, Jinlong},
  journal = {Physical Review B},
  volume = {98},
  issue = {15},
  pages = {155443},
  numpages = {7},
  year = {2018},
  month = {Oct},
  publisher = {American Physical Society},
  unused_doi = {10.1103/PhysRevB.98.155443},
  unused_url = {https://link.aps.org/doi/10.1103/PhysRevB.98.155443}
}

@article{Ma2021,
  author  = {Ma, Hai-Yang and Hu, Mengli and Li, Nana and Liu, Jianpeng and Yao, Wang and Jia, Jin-Feng and Liu, Junwei},
  title   = {Multifunctional Antiferromagnetic Materials with Giant Piezomagnetism and Noncollinear Spin Current},
  journal = {Nature Communications},
  volume  = {12},
  pages   = {2846},
  year    = {2021},
  unused_doi     = {10.1038/s41467-021-23127-7}
}

@article{PhysRevB.54.11169,
  title = {Efficient Iterative Schemes for Ab Initio Total-Energy Calculations Using a Plane-Wave Basis Set},
  author = {Kresse, G. and Furthm{\"u}ller, J.},
  journal = {Physical Review B},
  volume = {54},
  issue = {16},
  pages = {11169--11186},
  numpages = {0},
  year = {1996},
  month = {Oct},
  publisher = {American Physical Society},
  unused_doi = {10.1103/PhysRevB.54.11169},
  unused_url = {https://link.aps.org/doi/10.1103/PhysRevB.54.11169}
}

@article{KRESSE199615,
title = {Efficiency of Ab-Initio Total Energy Calculations for Metals and Semiconductors Using a Plane-Wave Basis Set},
journal = {Computational Materials Science},
volume = {6},
number = {1},
pages = {15-50},
year = {1996},
issn = {0927-0256},
unused_doi = {https://doi.org/10.1016/0927-0256(96)00008-0},
unused_url = {https://www.sciencedirect.com/science/article/pii/0927025696000080},
author = {G. Kresse and J. Furthmüller},
}

@article{PhysRevB.50.17953,
  title = {Projector Augmented-Wave Method},
  author = {Bl{\"o}chl, P. E.},
  journal = {Physical Review B},
  volume = {50},
  issue = {24},
  pages = {17953--17979},
  numpages = {0},
  year = {1994},
  month = {Dec},
  publisher = {American Physical Society},
  unused_doi = {10.1103/PhysRevB.50.17953},
  unused_url = {https://link.aps.org/doi/10.1103/PhysRevB.50.17953}
}

@article{PhysRevLett.77.3865,
  title = {Generalized Gradient Approximation Made Simple},
  author = {Perdew, John P. and Burke, Kieron and Ernzerhof, Matthias},
  journal = {Physical Review Letters},
  volume = {77},
  issue = {18},
  pages = {3865--3868},
  numpages = {0},
  year = {1996},
  month = {Oct},
  publisher = {American Physical Society},
  unused_doi = {10.1103/PhysRevLett.77.3865},
  unused_url = {https://link.aps.org/doi/10.1103/PhysRevLett.77.3865}
}

@article{MADSEN2018140,
title = {BoltzTraP2, a Program for Interpolating Band Structures and Calculating Semi-Classical Transport Coefficients},
journal = {Computer Physics Communications},
volume = {231},
pages = {140-145},
year = {2018},
issn = {0010-4655},
unused_doi = {https://doi.org/10.1016/j.cpc.2018.05.010},
unused_url = {https://www.sciencedirect.com/science/article/pii/S0010465518301632},
author = {Georg K. H. Madsen and Jesús Carrete and Matthieu J. Verstraete},
}

@article{xuexiongxiongPRB,
  title = {Multidimensional Control of Altermagnetism via Symmetry Engineering in van der Waals Heterostructures},
  author = {Chen, Xin and Yuan, Jingbiao and Meng, Haiyu and Fang, Shibo and Ang, Yee Sin and Xue, Xiong-Xiong and Chen, Ke-Qiu and Tang, Li-Ming},
  journal = {Physical Review B},
  volume = {113},
  issue = {17},
  pages = {174411},
  numpages = {10},
  year = {2026},
  month = {May},
  publisher = {American Physical Society},
  unused_doi = {10.1103/4n2d-bqtp},
  unused_url = {https://link.aps.org/doi/10.1103/4n2d-bqtp}
}

@article{Rondinelli2008,
  author  = {Rondinelli, James M. and Stengel, Massimiliano and Spaldin, Nicola A.},
  title   = {Carrier-Mediated Magnetoelectricity in Complex Oxide Heterostructures},
  journal = {Nature Nanotechnology},
  volume  = {3},
  number  = {1},
  pages   = {46--50},
  year    = {2008},
  unused_doi     = {10.1038/nnano.2007.412}
}

@article{Sui2021,
  author  = {Sui, Xuelei and Wang, Jianfeng and Yam, ChiYung and Huang, Bing},
  title   = {Two-Dimensional Magnetic Anionic Electrons in Electrides: Generation and Manipulation},
  journal = {Nano Letters},
  volume  = {21},
  number  = {9},
  pages   = {3813--3819},
  year    = {2021},
  unused_doi     = {10.1021/acs.nanolett.1c00172}
}
\end{document}